# Fast Bunch Integrators at Fermilab during Run II[*]

**Thomas Meyer**[a], **Charles Briegel**[a], **Brian Fellenz**[a] **and Greg Vogel**[a]

[a] *Fermi National Accelerator Laboratory,*
*Batavia, IL, U.S.A.*

*E-mail:* tsmeyer@fnal.gov

ABSTRACT: The Fast Bunch Integrator is a bunch intensity monitor designed around the measurements made from Resistive Wall Current Monitors. During the Run II period these were used in both Tevatron and Main Injector for single and multiple bunch intensity measurements. This paper presents an overview of the design and use of these systems during this period.

---

[*]Work supported by Fermi Research Alliance, LLC under Contract No. DE-AC02-07CH11359 with the United States Department of Energy.

## Contents

## 1. Introduction

The need for greater luminosity in high energy physics experiments demands ever increasing control and understanding of the particle beams used for them. Part of this understanding is a more detailed view of the beam. The Fast Bunch Integrator (FBI) is used at Fermilab to measure individual bunch intensities in both the Main Injector (MI) and the Tevatron (TeV) accelerators. This measurement became necessary during the Run II era as the push for higher intensity caused tune shifts in the counter-circulating Tevatron beams due to the focusing nature of proton bunches passing through oppositely charged anti-proton bunches. Since the dominant limit to the luminosity in the Tevatron is the antiproton production rate, improvements in luminosity come from smaller effects. The Main Injector FBI, for example, is used to measure the efficiency of coalescing multiple bunches into a single bunch. Higher efficiencies ultimately translate into higher luminosity.

## 2. Design Overview

The design of the Fast Bunch Integrator (FBI) is well documented in [1][2]. Since this paper will focus on the Run II uses of this system, a simple overview of the design will suffice. The FBI is based on signals from a Resistive Wall Current Monitor (RWCM) [3][4]. This is a beam line device that produces a waveform signal proportional in size and time for bunched beam traveling through the accelerator. The signal is then fed into an integrator board that resides in the VME chassis. The resulting voltage from the integrator is then digitized and read into the front-end, also residing in the VME chassis. In the front-end averaging and buffering is done to the numbers before the data is read out via the Accelerator Controls Network (ACNet). A simple diagram of the Main Injector FBI can be seen in Figure 1.

The Resistive Wall Current Monitor is AC coupled to the signal integrator board. This means that the FBI front-end must calculate and subtract a background value. This requires measuring the offset when beam is in the machine, but measured at the gaps between the beam bunches. Since, at these times, the system is reading a larger percentage of noise, the background measurement can become a noise source to the system measurements. Averaging of the background readings greatly reduces the noise of the data and makes for more stable measurements. This is most important in the Main Injector for anti-proton readings and in the Tevatron for all readings.

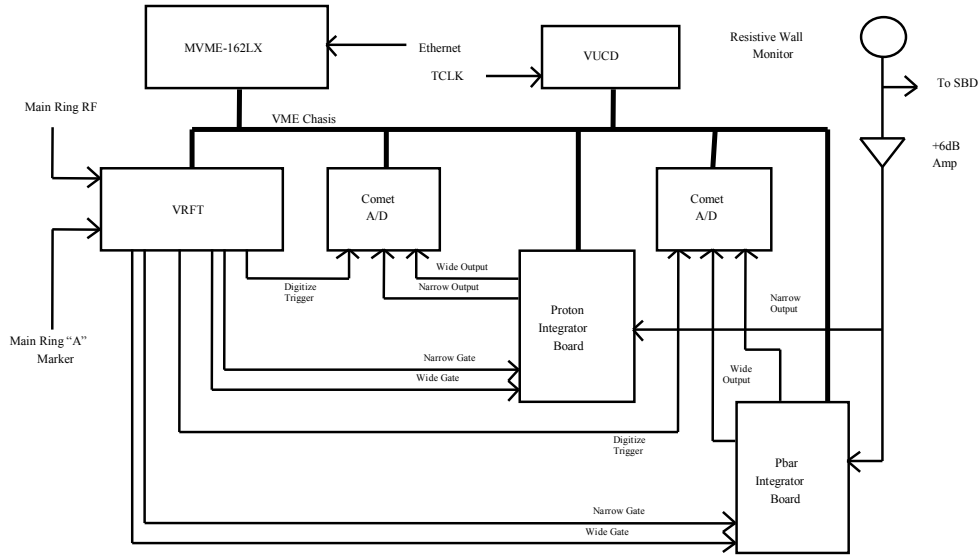

**Figure 1.** Block diagram of the original Fast Bunch Integrator design.

The FBIs make two basic forms of measurement, wide-gate and narrow-gate. In the wide-gate measurement the integrator is timed to integrate signal over multiple rf buckets. This allows it to read multiple beam bunches. The narrow-gate measurement is limited to one rf bucket (~19nsec). This insures that the measurement will only see one beam bunch for each measurement.

### 3. Measurements

#### 3.1 Main Injector

The FBI located in the Main Injector Ring was developed for and used exclusively for Colliding Beams physics. If used for the upcoming neutrino physics program it will require software changes that will reduce the current performance.

The Main Injector FBI makes measurements of three varieties. One Shot reads are made by general and logging applications. If this is the only mode running on the front-end, a set of measurements is taken, scaled, and returned to ACNet. This can happen as fast as 15 Hz and is the most common type of read requested from the users.

The second measurement type is the Fast Time Plot (FTP) measurement. In this measurement data is collected and scaled continuously at 720 Hz. The data is then transferred to the buffers, one for each FTP being requested via the control system. Data transfer to buffers and the subsequent transfer to ACNet occurs at a 15 Hz rate. This is the most common form of plotting used.

The Snapshot Plot is the third and fastest of the Main Injector FBI measurements. In this method the system free runs until the digitizer board's buffer is full. Necessarily this stops the other two methods of data collection during the Snapshot Plots collection, and is therefore used more sparingly than the other two methods. The high speed of a Snapshot Plot (23 KHz) makes it a desirable tool in accelerator performance analysis.

Anti-protons are a smaller signal than protons and much more susceptible to noise interference. Because of this noise problem the anti-proton reading takes 4 readings of each narrow and wide gate measurement. Four background measurements are also taken. The measurements are then averaged for each bunch and background. This has reduced noise by the expected factor of 2.

## 3.2 Tevatron

The Tevatron FBI, like the Main Injector also has the three methods of measurement, but they are achieved differently. In Tevatron the FBI front-end is collecting data at 360 Hz all the time. Each data collection occurs over 16 revolutions measuring each bunch 8 times. Each of the three gaps in beam is measured 16 times for a total of 48 background samples. This reduces noise greatly and allows the Tevatron's FBI measurement to be much more accurate. We are able to take a slower reading in the Tevatron because power supply and RF ramps do not change fast enough to require the faster Main Injector data rates. Once this data is collected the data is distributed to all the processes that have requests. This means that the Snapshot Plot has the same frequency as the Fast Time Plot. Again, this was a known trade off with the reduction of noise in the data.

The Tevatron also has a second FBI that is designed and used for broadcasting the proton and anti-proton intensities on the Machine Data (MDAT) link. MDAT is carried throughout the accelerator complex and contains many different signals capable of updating at 720 Hz. The data from the second FBIs is used by the automated collimator positioning system and will be discussed further in the next section.

## 4. Uses during Run II

### 4.1 Main Injector

The measurements of wide gate and narrow gate beam in Main Injector have made the FBI mission critical. Because of its speed the FBI can take measurements after acceleration has finished but before multi-batch to single-batch coalescing has occurred. Along with a set of measurements taken after coalescing, this information can give a quick and accurate measurement of the coalescing efficiency in Main Injector. This has become so useful that a task was added to the FBI software to enable or disable proton transfers into the Tevatron based both on the coalescing efficiency calculation and the single bunch measurements, both done by the FBI. The FBI's ability to measure bunch by bunch measurements in a quick fashion have made it irreplaceable when measuring independent efficiencies when transferring beam from either the Accumulator or the Recycler rings.

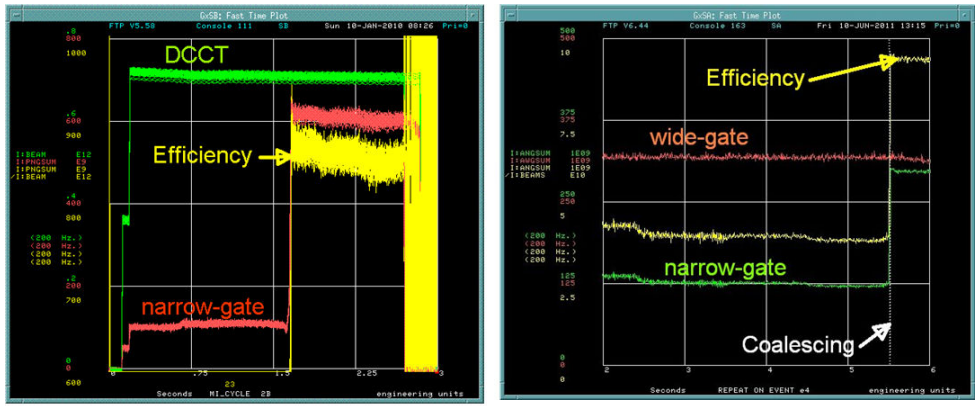

**Figure 2.** In the left plot, proton coalescing can be seen first in the red narrow gate trace and then from the yellow efficiency trace of the narrow gate divided by the total DCCT beam. In the right plot, anti-proton coalescing can be seen with the wide gate in red, the narrow in green and the coalescing efficiency in yellow.

**4.2 Tevatron**

In the Tevatron the FBI measurements are first compared to the Main Injector measurements to calculate bunch by bunch transfer efficiencies. A comparison of wide gate measurements to narrow also is used to measure the size of satellite bunches that can cause beam loss later in the proton-antiproton colliding beams. While the beams are colliding, the bunch by bunch readings are used to calculate the lifetime of proton and anti-proton beams independently.

Once the beams have reached their peak energy, they still undergo a process of "scraping" to remove excess particles outside the central core. This excess beam, called "halo", can cause high losses in the experimental detectors that can obscure data collection and damage sensitive detectors that lie within millimetres of the circulating beam. Originally control of the scrapers was done manually. When an automated solution was designed it required a realtime measurement of both proton and anti-proton intensities that could be made independently. A second FBI was developed for this purpose. Its sole function is to measure the proton and anti-proton intensities and broadcast them on the Machine Data (MDAT) link in the accelerator controls system. This link carries many types of data on frames at 720 Hz around to all accelerators and can be used for a variety of purposes. The data from the second FBI is used to ensure the collimators do not scrape away the core of the proton or anti-proton beam.

**5. Conclusions**

During the Run II era the Fast Bunch integrators have found a multitude of uses. From anti-proton transfers to muti-bunch beam coalescing, Main Injector transfers to halo scraping and lifetime measurements, the Fast Bunch Integrators have proved invaluable in the creation and maintenance of Colliding Beams stores at Fermilab.

**Acknowledgments**